\documentclass[useAMS, usenatbib]{mnras}
\usepackage{times}
\usepackage{amssymb}
\usepackage{amsmath}
\usepackage{bm}
\usepackage{rotating}
\usepackage{epsfig}
\usepackage{graphicx}
\input{epsf}

\title[Gravitational Redshifted Line Search In Ser X-1]
{Search For Gravitational Redshifted Absorption Lines In LMXB Serpens X-1}

\author[H. Yoneda et al.]
{Hiroki Yoneda$^{1,2}$, Chris Done$^{3}$, Frits Paerels$^{4}$, Tadayuki Takahashi$^{2,1}$, Shin Watanabe$^{2,1}$\\
\\
$^1$ Department of Physics, University of Tokyo, 7-3-1 Hongo, Bunkyo-ku, Tokyo, 113-0033, Japan \\
$^2$ Institute of Space and Astronautical Science, JAXA, 3-1-1 Yoshinodai, Chuo-ku, Sagamihara, Kanagawa, 252-5210, Japan \\
$^3$ Department of Physics, University of Durham, South Road, Durham, DH1 3LE, UK.\\
$^4$ Columbia Astrophysics Laboratory and Department of Astronomy, Columbia University, 538 W. 120th St., New York, NY 10027, USA \\
}

\date{Submitted to MNRAS}
\pagerange{\pageref{firstpage}--\pageref{lastpage}} \pubyear{}

\def\apj{ApJ}

\begin{document}

\topmargin = -0.5cm

\maketitle

\label{firstpage}

\begin{abstract}
The equation of state for ultra-dense matter can be tested from observations of the ratio of mass to radius of neutron stars. 
This could be measured precisely from the redshift of a narrow line produced on the surface. 
X-rays bursts have been intensively searched for such features, but so far without detection. 
Here instead we search for redshifted lines in the persistent emission, where the accretion flow dominates over the surface emission. 
We discuss the requirements for narrow lines to be produced, 
and show that narrow absorption lines from highly ionized iron can potentially be observable in accreting low mass X-ray binaries (low B field) which have either low spin or low inclination so that Doppler broadening is small. This selects Serpens X-1 as the only potential candidate persistent LMXB due to its low inclination. 
Including surface models in the broad band accretion flow model predicts that the absorption line from He-like iron at 6.7~keV should be redshifted to $\sim 5.1~\mathchar`-~5.7$~keV (10 - 15~km for $1.4\mathrm{M_\odot}$)
and have an equivalent width of $0.8~\mathchar`-~8$~eV for surface temperatures of $7~\mathchar`-~10\times 10^6$~K. 
We use the high resolution {\it Chandra} grating data to give a firm upper limit of $2~\mathchar`-~3$ eV for an absorption line at $\sim 5$~keV. 
We discuss possible reasons for this lack of detection (the surface temperature and the geometry of the boundary layer etc.). 
Future instruments with better sensitivity are required in order to explore the existence of such features.
\end{abstract}


%==============================================
\section{Introduction} 
\label{sec:introduction}
%==============================================

Neutron star masses and radii are determined by the equation of state
(EoS) of dense matter, described by quantum chromodynamics (QCD) at
the quark-gluon interaction level. However, this is not a theory
which is well understood. Only small portions of the two dimensional
phase space (temperature and chemical potential or equivalently
pressure) are satisfactorily described by theory and accessible to
experiment. Neutron star cores have mean densities which are much
higher than can be produced in current laboratory conditions, 2 - 8
times larger than the nuclear saturation density. They are also
relatively cool and in equilibrium with both strong and weak force
(unlike heavy ion collider experiments) and are neutron rich (unlike
normal nuclei which are approximately symmetric in neutrons and
protons). Thus measuring the macroscopic properties of neutron stars 
gives insight into the fundamental QCD interactions in a new regime
(e.g. \citealt{Lattimer:2012}).

The most stringent constraints so far come from the firm detection of
neutron stars with masses of $\sim 2 M_\odot$ \citep{Demorest:2010,Antoniadis:2013}, 
directly ruling out any EoS which has a
maximum mass below this value. This was first thought to exclude a
significant contribution of hyperons \citep{Demorest:2010}, which is
puzzling as they are energetically favorable at high
densities. However, more recent calculations of the effect of hyperons
show that these were not necessarily inconsistent with the data
(e.g. \citealt{Whittenbury:2014}). 

Ideally, the full EoS can be traced out from measuring both mass and
radius for a sample of neutron stars of different masses. Some of the
best current constraints for this come from thermal emission from the
surface of quiescent low mass X-ray binaries (LMXB), the cooling
tails of thermonuclear bursts and pulse profile modelling of
accreting millisecond pulsars (e.g the review by \citealt{Ozel:2016}). However, there are multiple caveats for each technique (see
e.g. \citealt{Miller:2016}), and even the best determinations have
uncertainties of the order of 10 - 20\%.

These uncertainties could be reduced by an order of magnitude through an unambiguous measure of $M/R$ from the surface redshift of a narrow atomic line. 
The line is redshifted by the strong gravity of the neutron star 
and the redshift parameter $z$ is connected with the ratio of mass to radius by general relativity:
\begin{eqnarray*}
  1 + z = \left(1 - \frac{2GM}{c^2 R} \right)^{-0.5}
\end{eqnarray*}
Emission from the neutron star surface dominates during X-ray bursts,
so these have been extensively studied. 
\cite{Cottam:2002} reported the detection of Fe XXVI/XXV and O VIII absorption lines
with $z=0.35$ from the spectra of X-ray bursts from EXO 0748-676, but
these features were not reproduced in more sensitive data \citep{Cottam:2008}. 
Subsequent determination of a high spin for this
object further showed that these narrow features cannot be produced from the
surface \citep{Galloway:2010}. 
No other narrow line features have been convincingly detected to date.

We review the requirements for such absorption line features to be
produced, and show that the only feasible persistent source where these
might be detected is the LMXB Serpens X-1 (Section~\ref{sec:requirements_for_detection}). We describe the
expected features from models of the surface (Section~\ref{sec:nsmodel}), and use
these models combined with the {\it Suzaku} broadband data to predict the
equivalent width of the most prominent absorption line, Fe XXV (Section~\ref{sec:suzaku_data_analysis}). 
Section~\ref{sec:chandra_data_analysis} shows that these predictions are already challenged by
upper limits on this feature from {\it Chandra} grating data for a high
temperature surface. We discuss physical implications of our
results on the thermal conductivity of neutron stars and other physical parameters, and conclude by
summarizing our results.

\section{Requirements for observable narrow absorption lines from the
neutron star surface}
\label{sec:requirements_for_detection}

We follow the discussion in the Astro-H white paper on Low-mass X-ray Binaries \citep{WP:2014}. 
To see narrow absorption lines from the NS surface requires that there
are heavy elements in the photosphere, that these are not completely
ionized, that the photosphere is not buried beneath an optically thick
accretion flow, and that the resulting atomic features are not
substantially broadened.

Heavy elements are deposited onto the NS surface by the accretion
flow. They are stopped by collisional processes, which are more
efficient for higher mass/charge ions. Hence iron and other heavy
elements are halted higher up in the photosphere than lower atomic
number elements. They can then be destroyed by spallation bombardment
(by the still energetic helium and hydrogen ions, transforming the
iron nuclei to lower Z elements) or sink under gravity. The
deposition and destruction rate both depend linearly on $\dot{M}$ so
the steady state Fe column is around the solar abundance, independent of $\dot{M}$
for $L_\mathrm{x}>6\times 10^{32}$~ergs/s
\citep{Bildsten:2003,Chang:2005}. Thus there can be iron and other
heavy elements in the photosphere of an accreting NS, but not in an
isolated or very quiescent neutron star.

Accreting neutron stars can have either low or high mass companion
stars. The neutron stars in high mass X-ray binaries are young, and the neutron stars
typically have very high magnetic fields. These broaden any potential
atomic features via Zeeman splitting, with $\Delta E\sim
12~B/(10^{9}~\mathrm{G})$~eV \citep{Loeb:2003}. 
By contrast, the LMXB typically have 
low fields with $B \leq 10^9~\mathrm{G}$, so narrow atomic features can possibly form in these systems.

\begin{figure}
\begin{center}
\includegraphics[bb = 0 0 2000 400, width=8cm]{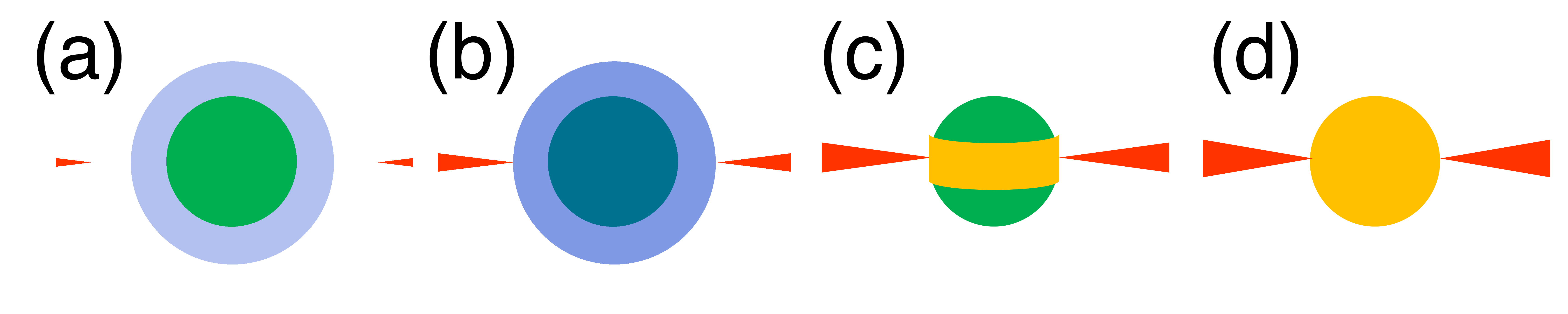}
\caption{The geometry of the accretion flow onto the NS surface depending on the accretion rate. $L / L_\mathrm{Edd}$ is $\sim 10^{-4}~(a),\sim10^{-2}~(b),\sim 0.1~(c),\sim 1~(d)$ from left to right \citep{WP:2014}.}
\label{fig:NStype}
\end{center}
\end{figure}

\begin{figure}
\begin{center}
\includegraphics[bb = 0 0 1182 632, width=8.5cm]{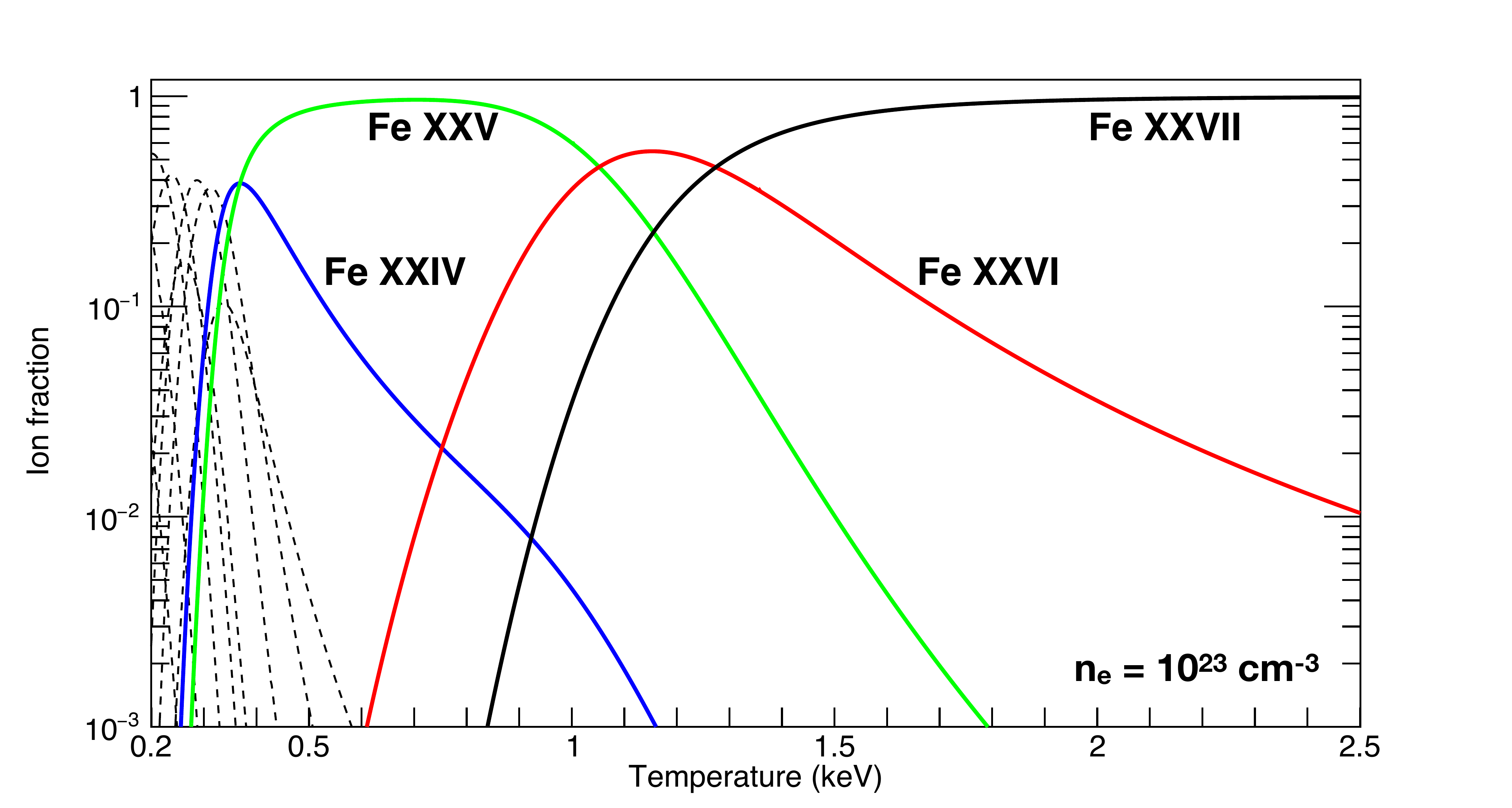}
\caption{Ion fraction of Fe in the neutron star atmosphere. The electron density is fixed to $10^{23}~\mathrm{cm^{-3}}$ which is the typical value in the NS atmosphere \citep{Ozel:2013}. It is calculated using the Saha equation.
We include the effect of pressure ionization by lowering the ionization potential due to Debye shielding. }
\label{fig:ion}
\end{center}
\end{figure}

The surface of a neutron star in an LMXB can only be seen if it is not
hidden beneath the accretion flow. This depends on the geometry of
the accreting material as well as its optical depth. In the truncated
disc/hot accretion flow models, the accretion geometry changes
dramatically at the spectral transition between the island and banana
branches (see e.g. \citealt{Done:2007, Kajava:2014}). 
At low accretion rates, the accretion flow is hot and
quasi-spherical interior to some truncation radius at which the thin
disc evaporates (island state). There is additional luminosity from
the boundary layer where the flow settles onto the surface, but the flow
and boundary layer merge together, forming a single hot ($\sim
30~\mathchar`-~50$~keV), optically thin(ish) ($\tau\sim 1.5~\mathchar`-~2$) structure
\citep{Medvedev:2001}. A fraction $e^{-\tau}$ of the surface emission
should escape without scattering, so $\sim 10~\mathchar`-~20$\% of the intrinsic
NS photosphere should be seen directly (see Figure~\ref{fig:NStype} a
and b). The temperature of this surface emission can also be seen
imprinted onto the low energy rollover of the Compton spectrum and is
only $\sim 0.5~\mathchar`-~0.6$~keV (e.g. \citealt{Sakurai:2014}). 
Figure~\ref{fig:ion} shows the ion fraction of Fe in the neutron star atmosphere 
assuming local thermodynamic equilibrium.
At $0.5~\mathchar`-~0.6$~keV, there should be a considerable fraction of iron which is
not completely ionized, so atomic features could be seen
\citep{Rauch:2008}, though X-ray irradiation from the optically thin
boundary layer could give a more complex photosphere temperature
structure. 

At higher mass accretion rates ($L\ge 0.1 ~\mathchar`-~ 0.5~L_\mathrm{Edd}$), the thin
disc extends down to the NS surface, forming a boundary layer where it
impacts around the NS equator. The boundary layer is now optically
thick ($\tau\sim 5~\mathchar`-~10$) so hides the surface beneath it. The boundary
layer itself is at the local Eddington temperature of $\sim 2.5$~keV
\citep{Revnivtsev:2013}. This is high enough that iron should be
almost completely ionized (Figure~\ref{fig:ion} and \citealt{Rauch:2008}), so no atomic features
are expected from the luminous accretion flow. However, the vertical
extent of the boundary layer depends on the accretion rate, and it
only forms an equatorial belt for $L\lesssim 0.3L_\mathrm{Edd}$
(\citealt{Suleimanov:2006}, Figure~\ref{fig:NStype} c). The pole is
uncovered, so this part of the neutron star surface can be seen
directly. It is heated mainly by thermal conduction from the
equatorial accretion belt, so its temperature is probably cool enough
for H- and He-like iron to exist.

At still higher mass accretion rates, the spreading layer extends up
to the pole and the surface is completely covered by the optically
thick, completely ionized accretion flow (upper banana branch and Z
sources: Figure~\ref{fig:NStype} d). Hence the largest fraction of
surface emission should be seen from a pole-on view of a lower banana
branch source, where the optically thick accretion flow is confined to
an equatorial belt, or in an island state, where the accretion flow
covers most of the surface but is optically thin (e.g. \citealt{Sakurai:2014}).

The final requirement is that the atomic features are not broadened by
rotation, where $\Delta E \sim
1600~\nu_\mathrm{spin}/(600~\mathrm{Hz}) \sin i$~eV \citep{Ozel:2013}.
This is a stringent constraint as typical LMXBs have
$\nu_\mathrm{spin}=185~\mathchar`-~650$~Hz \citep{Altamirano:2012, Patruno:2012}.

There is one system, Serpens X-1, where optical spectroscopy of the 2
hour binary orbit indicates a low inclination, $i\le 10^\circ$
\citep{Cornelisse:2013}. A low inclination is also consistent with
the non-detection of dips in the X-ray lightcurves, and the lack of
any burst oscillations in the X-ray burst lightcurves
\citep{Galloway:2008}. This persistent system is always in the soft
state (mid banana branch, $L/L_\mathrm{Edd} \sim 0.5$
\citep{Chiang:2016}, so the boundary layer should not extend over the
pole. Additionally, this is a very bright source $\sim$ 300 mCrab.
Thus Serpens X-1 is the only currently known persistent source where
it may be possible to detect gravitationally redshifted lines from the
surface during normal (non-burst) accretion.

%==============================================
\section{Neutron star surface blackbody model}
\label{sec:nsmodel}
%==============================================

The emission from the hot neutron star surface should contain atomic
features, which depend on the temperature, gravity and chemical
composition of the photosphere. \cite{Rauch:2008} show calculations
for a neutron star of a mass of 1.4 M$_{\odot}$ and a radius of 10~km (i.e. $\log
g = 14.39$, redshift parameter $z = 0.306$) and the solar abundance over the
temperature range $1~\mathchar`-~20\times 10^6$~K ($\sim$ 0.1 - 2~keV). The focus
of their work was on the claimed detection of iron $n =3~\mathchar`-~2$ absorption lines
in EXO 0748-676 \citep{Cottam:2002}, but they also show the Fe $n = 2~\mathchar`-~1$
absorption lines which are in a simpler part of the spectrum where
there is less ambiguity in interpretation. These show that 
these H- and He-like K shell absorption lines are
strongest around surface temperatures
of 1~keV as iron is mostly ionized for temperatures above 2~keV (Figure~\ref{fig:ion}).
When the temperature is below 0.3 keV,
the luminosity of the surface blackbody is (0.3/1)$^4$ i.e. 100 times weaker
so that it is difficult to observe the lines.

\begin{figure}
\begin{center}
\includegraphics[bb = 0 0 750 1000, width=80mm]{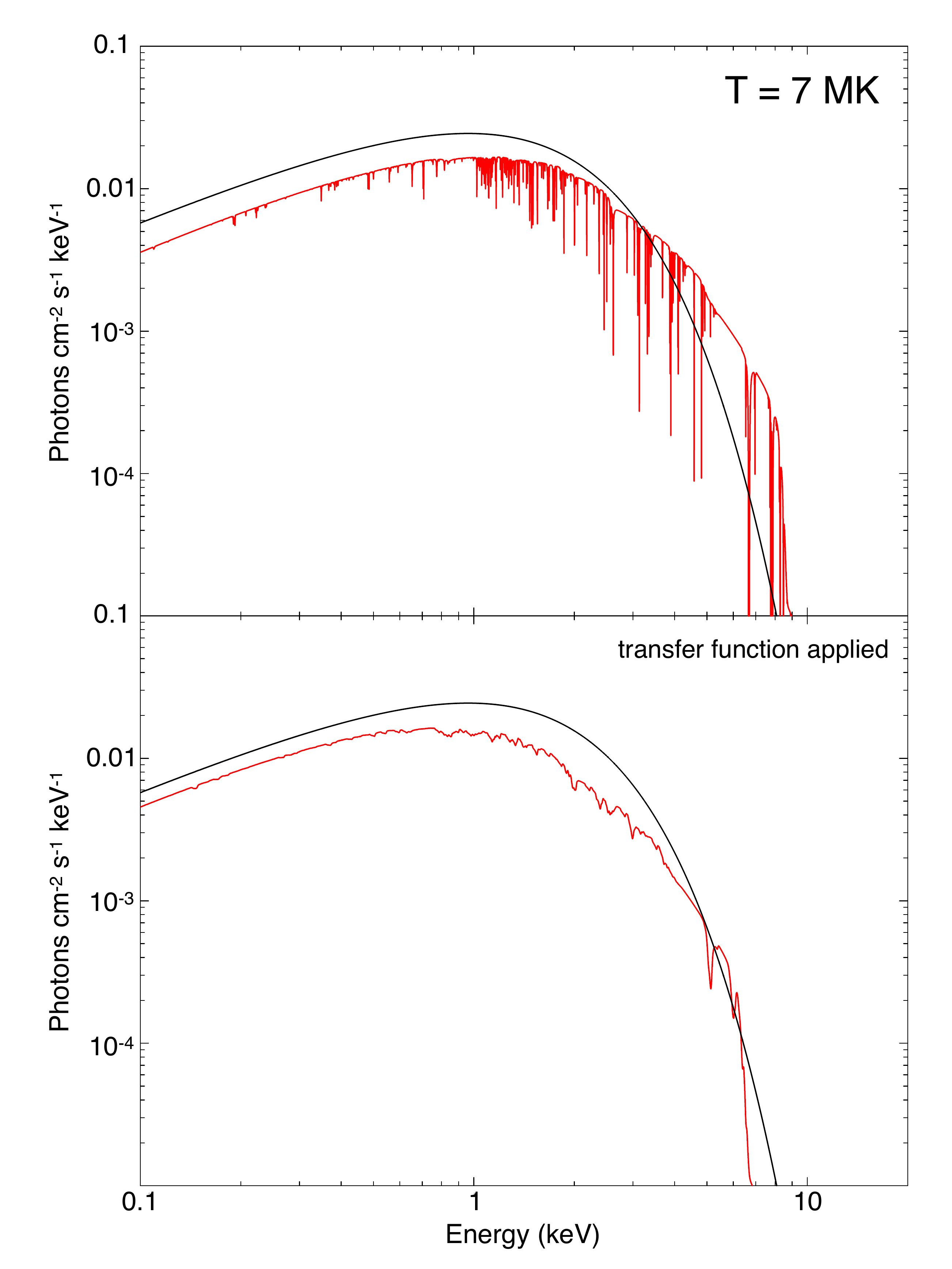}
\includegraphics[bb = 0 0 750 1000, width=80mm]{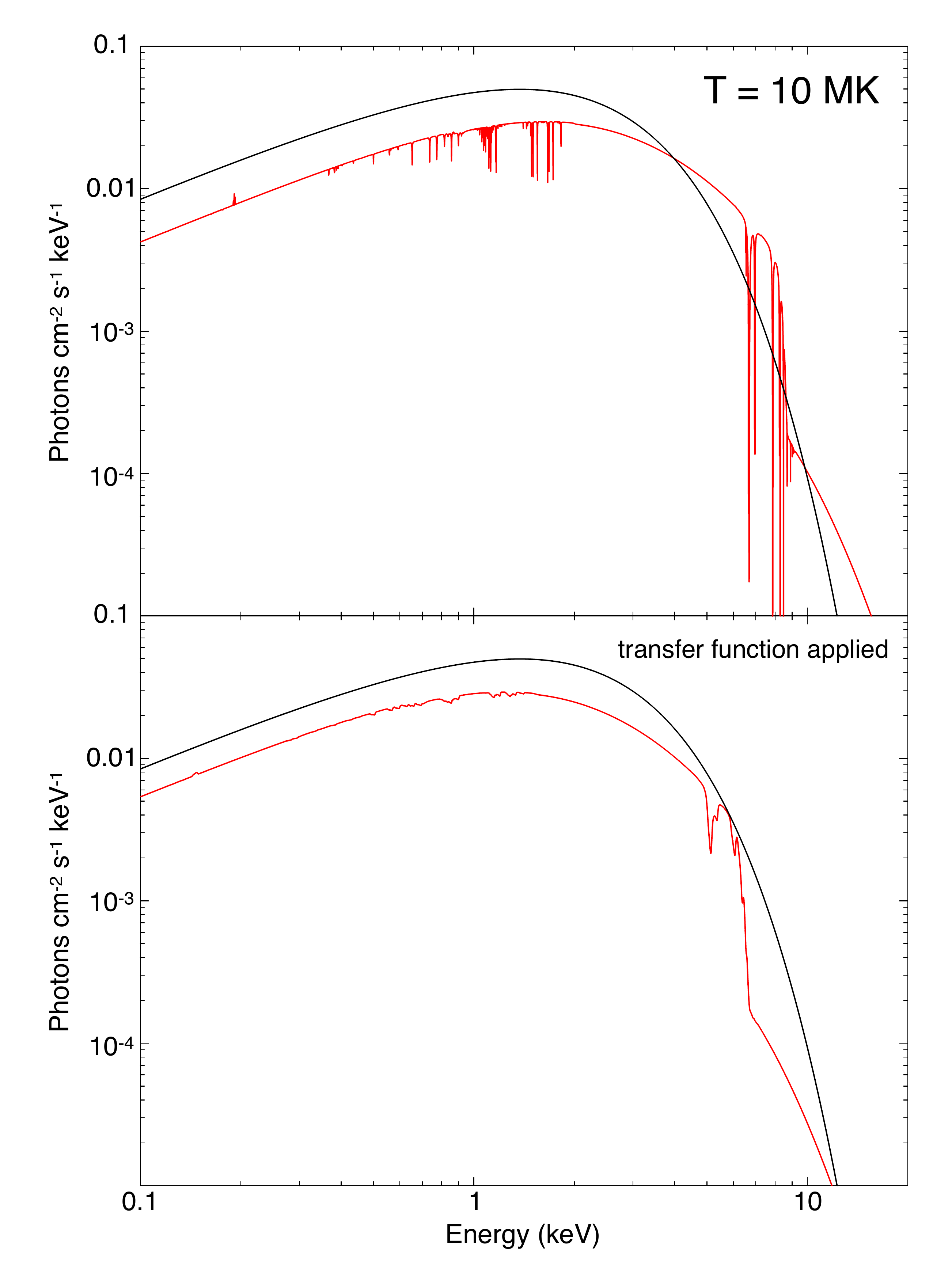}
\caption{Neutron star atmosphere models with T = $7,10\times10^6$ K. The red curve in the top is the spectrum without the redshift and the broadening effect. The bottom is the spectrum after the transfer function is applied. The black curves are the blackbody spectra with T = $7,10\times10^6$ K.}
\label{fig:nsbbmodel}
\end{center}
\end{figure}

The neutron star surface temperature can be seen in neutron stars at
low mass accretion rates (island state) as it is imprinted on the
Comptonized boundary layer emission as the seed photon temperature,
and distinctly hotter than the disc photons. \cite{Sakurai:2014} show
that this is $\sim 0.5$~keV for Aql X-1 in the brighter island states,
and similar seed photon temperatures ($0.5~\mathchar`-~0.7$~keV) are seen in
other neutron stars in similar states \citep{Gierliski:2002,DiSalvo:2015}. 
Higher temperatures of $1$ - $1.6$~keV are seen (again
from seed photons) in the higher mass accretion rate (banana branch)
states \citep{Oosterbroek:2001,Gierliski:2002,Sakurai:2014,DiSalvo:2015},
though these are for the heated surface
underneath the boundary layer rather than measuring the temperature at
the uncovered pole where the temperature should be somewhat
lower. Hence we use two models for the neutron star 
with effective temperatures of $7\times 10^6$ and $10
\times$ $10^6$~K (Figure~\ref{fig:nsbbmodel}) to explore their
predictions for the visibility of the iron absorption line from the
surface. These are calculated as in \cite{Rauch:2008} 
(Suleimanov, private communication). The spectra are given as
intrinsic (unredshifted) surface Eddington flux $H(E)=F(E)/4\pi$ so we
convert these to surface luminosity $L(E)=(4\pi R)^2 H(E)$.

We predict their spectra at infinity using the calculated relativistic
transfer functions of \cite{Baubock:2013}. These include the Doppler
effects from spin as well as gravity, so they depend on inclination
and spin frequency as well as surface gravity. We assume the system
inclination of $10^\circ$ together with a typical neutron star spin
frequency of 400Hz for the same $\log g$ as above, so $z=0.307$, and
convolve the neutron star spectra with this transfer function
(Baub{\"o}ck, private communication).

%==============================================
\section{Estimation of the absorption line intensity with {\it Suzaku}} 
\label{sec:suzaku_data_analysis}
%==============================================
\subsection{Observations and data reduction} 

{\it Suzaku} observed Serpens X-1 for $\sim 30$~ks on 2006 Oct 24
(Obs. ID : 401048010). The observation of XIS/{\it Suzaku} was taken
in $\frac{1}{4}$ window with 1 s burst clock mode at the XIS nominal
pointing position. We correct the images by using {\tt aeattcor2} in
the HEAsoft package. The CCD data are still affected by pileup
\citep{yamada:2012}, so we remove the data
from a circle of 
radius 60 pixels centered on the brightest pixel, corresponding to a 3\%
pileup fraction. We sum XIS0, XIS2 and XIS3 using {\tt addascaspec} in the HEAsoft package and reprocessed HXD PIN
data following the standard analysis procedure and adopted {\tt
 ae\_hxd\_pinxinome3\_20080129.rsp} as the response file. 
The non-X-ray background is estimated following the standard analysis thread described on {\it The Suzaku Data Reduction Guide} and the cosmic X-ray background is ignored because the source is very bright.
The spectra were rebinned so that each bins contain more than 20 counts.

We used XSPEC version 12.9.0 to fit the spectra. The fitting was
performed from 1.0 keV to 9.0 keV in the XIS and from 15.0 keV to 20.0
keV in the HXD PIN. The region from 1.5 keV to 2.5 keV was ignored
due to the large calibration uncertainties coming from the
instrumental edges. 
We set the normalization of PIN data relative to XIS data as a free parameter.
The best fit values of it are consistent with the value (1.16) on {\it The Suzaku Data Reduction Guide} in the $\sim 1 \sigma$ confidence level (Table~\ref{table:nsbl}, \ref{table:nsbl_nsbb}).

We assumed a neutral hydrogen column density as $4.4 \times 10^{21}
\mathrm{cm^{-2}}$ \citep{Dickey:1990} via the {\tt tbnew\_gas} model
which is a new and improved version of the X-ray absorption model {\tt
 tbabs} \citep{Wilms:2000} and set the photoelectric absorption
cross-sections as "vern" and the metal abundances as "wilm". Errors
in this paper is given at 90\% confidence level unless otherwise
stated.

\subsection{Identification of the spectral components}
\label{subsec:spectralcomponent}

We fit the spectrum with a continuum model consisting of a disc and
Comptonized boundary layer. There is considerable spectral degeneracy
in decomposing a broadly curving continuum into two smoothly curving
components (e.g. \citealt{Done:2002,Revnivtsev:2006}), 
so we use the best physical models for the emission. 
We describe the disc by the {\tt kerrbb}
model \citep{Li:2005} which includes relativistic smearing of the sum
of color temperature corrected blackbody
components with luminosity given by the fully relativistic emissivity.
We fix the inclination angle of $10^\circ$ \citep{Cornelisse:2013} and the mass of 1.4~$\mathrm{M_\odot}$ and assume a distance of 10 kpc 
(the $X=0$ limit from type 1 X-ray burst observations; \citealt{Galloway:2008}).
We describe the Comptonisation using the {\tt nthcomp} model
\citep{Zdziarski:1996}. This should produce a reflection component
from illumination of the disc, but this can be complex
\citep{Bhattacharyya:2007, Cackett:2008, Miller:2013, Chiang:2016-2, Chiang:2016, Matranga:2017}. 
Here our focus is to describe the
continuum shape for a simulation of the narrow redshifted absorption
lines, so we simply use a Gaussian emission line to fit the data. The
resulting parameters are shown in Table~\ref{table:nsbl}. 

While the temperature of electrons in the boundary layer,
$T_e~(> 3.1\;\mathrm{keV})$ is
poorly constrained due to the lack of data in high energy band,
the temperature of the surface beneath it,
$T_{\mathrm{BL}}~(= 1.76_{-0.14}^{+0.03}\;\mathrm{keV})$ are close to those
derived for the boundary layer in similarly bright LMXB spectra by
\cite{Revnivtsev:2006}, where they break the spectral
degeneracies by Fourier resolved spectroscopy. Thus our spectral
decomposition appears reasonable.

The luminosity ratio of the disc emission to the boundary layer
emission is $0.943_{-0.029}^{+0.022}$. This potentially already
provides some constraints on the EoS modulo neutron star spin
\citep{Sibgatullin:2000}. The boundary layer dissipates the remaining
kinetic power of the accretion flow, which will depend on the
difference in spin frequency between the disc inner edge and the
surface. However, it also depends on the EoS, especially where the
neutron star is smaller than the radius of the last stable circular
orbit as the boundary layer luminosity is enhanced by the additional
kinetic energy of the radially plunging material. The calculations of
\cite{Sibgatullin:2000} (their Fig 1) give a spin of 540~Hz for
a luminosity ratio of 0.943 with their assumed EoS
($R=11.3$~km for a $1.4~M_\odot$ neutron star). 
This is a
reasonable spin frequency, but it is probably overestimated as our
models for the boundary layer luminosity neglected the reflection continuum.

The expected efficiency of accretion is $\sim 0.14$ (Fig 1 of
\citealt{Sibgatullin:2000}). Combining this with the mass accretion rate
through the disc of $0.64\times10^{18}\;{\mathrm{g/s}}$
(Table~\ref{table:nsbl}) gives a total luminosity which is $\sim 45\%$
of the Eddington mass accretion rate for a $1.4~M_\odot$ neutron
star. The total absorption corrected bolometric flux from the data is
instead $1.0 \times 10^{-8}\;\mathrm{ergs/cm^{2}/s}$, giving a
luminosity of $1.2 \times 10^{38}\;\mathrm{ergs/s}$ for the assumed
distance of 10 kpc, which is $\sim 70\%$ of the the Eddington
luminosity. This discrepancy is less than a factor 2, but could
indicate either that the distance is $7.7$~kpc, as derived from
assuming the solar abundance for the X-ray bursts \citep{Galloway:2008},
or that the EoS is different to that assumed in \cite{Sibgatullin:2000}.

\begin{table}
\begin{center}
\begin{tabular}{ccc}
\hline \hline
Componet & Parameter & \\
\hline
{\tt tbnew\_gas} & $N_\mathrm{H} \; (10^{22} \; \mathrm{cm^{-2}})$ & $(0.44)$\\
\hline
{\tt kerrbb} 
& $\eta$ & (0.0)\\
& a & (0.2)\\    
& $i$ (deg) & (10.0)\\
& $M_\mathrm{bh}$ ($\mathrm{M_\odot}$) & (1.4)\\ 
& $M_\mathrm{dd}$ ($\mathrm{10^{18}\;g/s}$) & $0.637_{-0.009}^{+0.010}$\\
& $D_\mathrm{bh}$ (kpc) & (10.0)\\
& hd & $1.653_{-0.019}^{+0.019}$\\
& rflag & (1.0)\\
& lflag & (0.0)\\
& norm & (1.0)\\
\hline
{\tt nthComp} 
& Gamma & $6.3_{-3.3}^{+0.3}$\\
& $kT_\mathrm{e}$ (keV) & $> 3.1$\\
& $kT_\mathrm{bb}$ (keV) & $1.76_{-0.14}^{+0.03}$\\
& $L_\mathrm{disc} / L_\mathrm{BL}$ & $0.943_{-0.029}^{+0.022}$ \\
\hline
{\tt Gaussian}
& Line E (keV) &$6.656_{-0.031}^{+0.031}$\\
& $\sigma$ (keV) & $0.233_{-0.036}^{+0.040}$\\
& norm ($10^{-3}$)& $2.58_{-0.33}^{+0.36}$\\
\hline
{\tt constant} (HXD PIN)
& & $1.205_{-0.044}^{+0.045}$\\
\hline
$\chi^2/d.o.f.$ & & 2097.5 / 1918\\
\hline
\hline
\end{tabular}  
\caption[t]{Best-fit parameters for {\tt kerrbb + nthcomp} without the surface blackbody models using {\it Suzaku} spectrum}
\label{table:nsbl}
\end{center}
\end{table}

\begin{figure}
\begin{center}
\includegraphics[bb = 0 0 750 600, width=85mm]{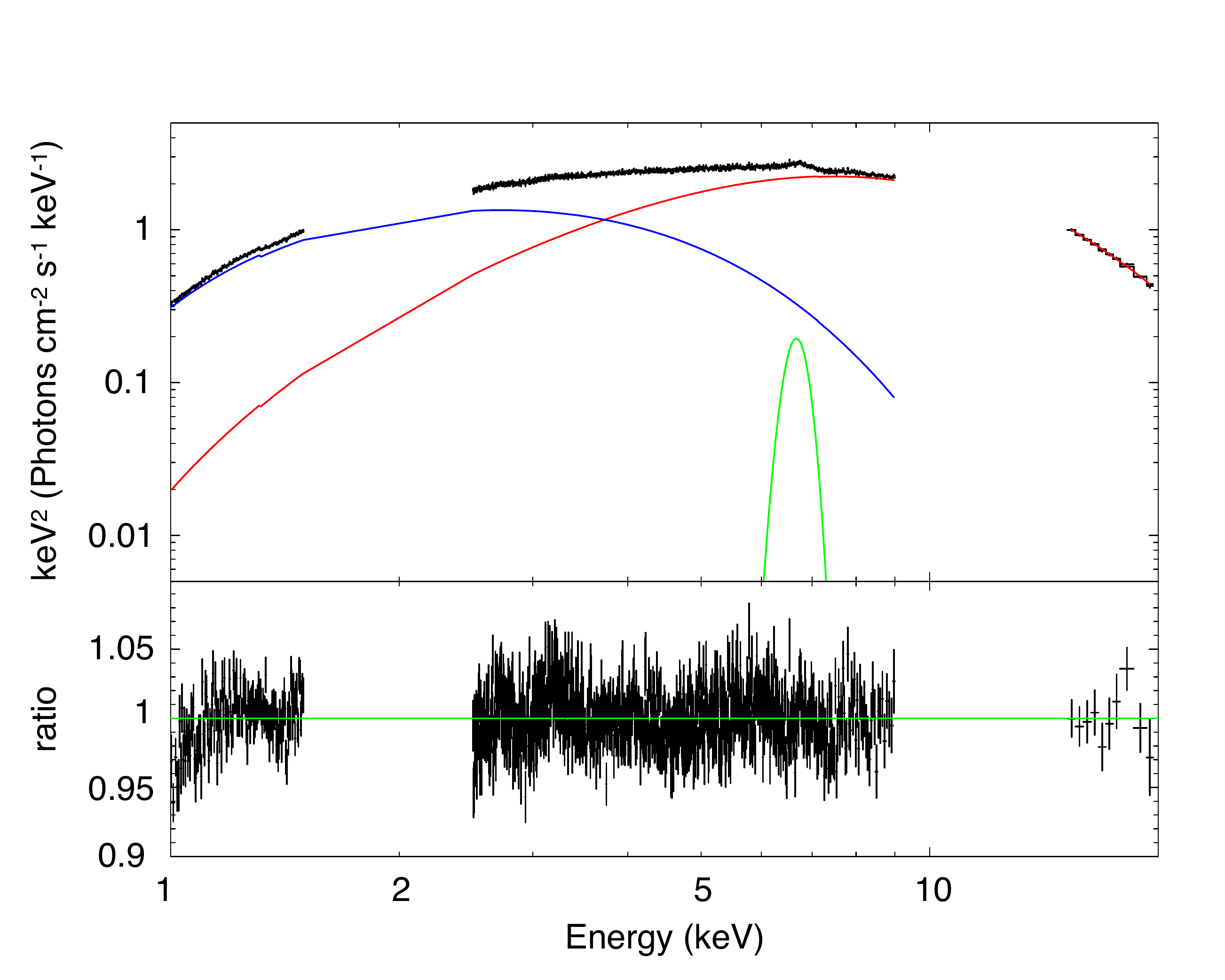}
\caption{Unfolded {\it Suzaku} spectrum with {\tt kerrbb + nthcomp} without the surface blackbody models. The blue curve is {\tt kerrbb}, the red one is {\tt nthcomp} and the green one is {\tt Gaussian} for the iron line.}
\label{fig:nsbl}
\end{center}
\end{figure}

\subsection{Estimation of the equivalent width of the iron absorption line}

\begin{figure*}
\begin{center}
\includegraphics[bb = 0 0 750 600,width=85mm]{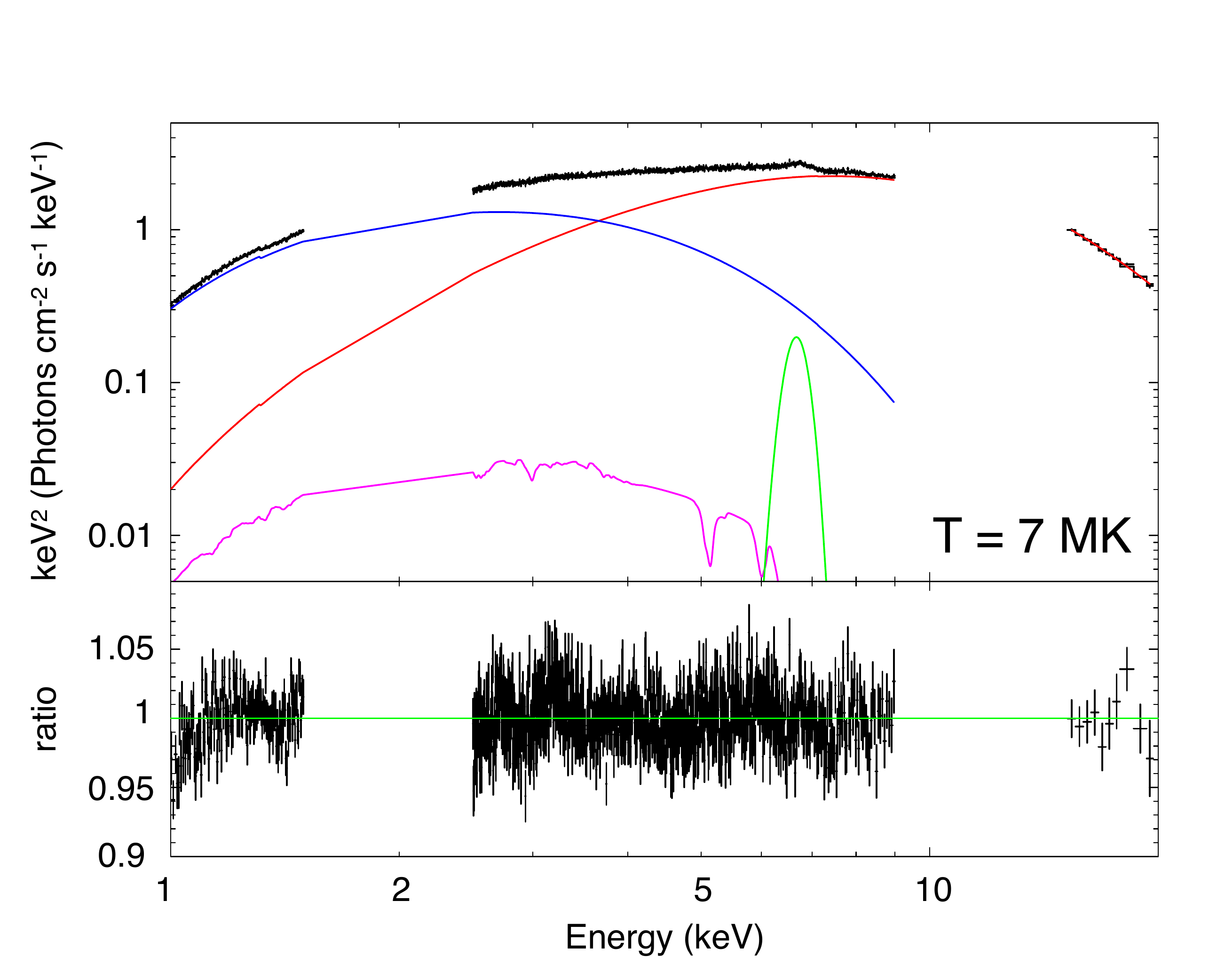}
\includegraphics[bb = 0 0 750 600,width=85mm]{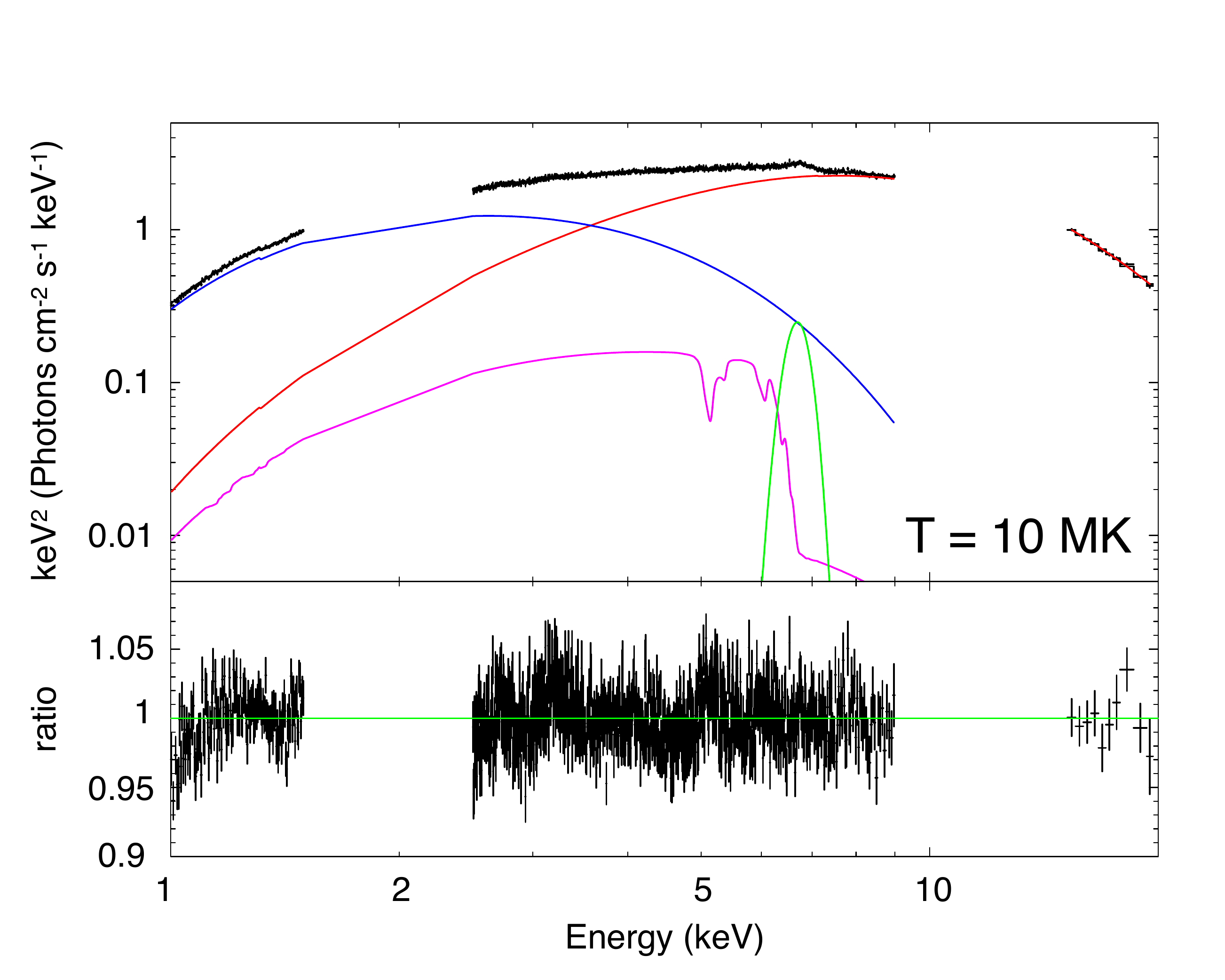}
\caption{Unfolded {\it Suzaku} spectrum with {\tt kerrbb + nthcomp} with the surface blackbody models. The magenta is the surface model and the blue curve is {\tt kerrbb}, the red one is {\tt nthcomp}, the green one is {\tt Gaussian} for the iron line. The effective temperature of the surface are $7\times10^6~\mathrm{K}$ in the left figure and $10\times10^6~\mathrm{K}$ in the right one.}
\label{fig:nsbl_nsbb}
\end{center}
\end{figure*}

We add each of the two different temperature surface spectra described
in Section~\ref{sec:nsmodel} to the emission model derived above. This is the maximum
possible contribution from the surface, as it assumes that the entire
star is directly visible rather than being partly covered by the
boundary layer and partly obscured by the disc. However, these effects
are minimized for a pole on view so this represents a reasonable
contribution of the surface emission for Serpens X-1. The results are
shown in Table~\ref{table:nsbl_nsbb} and Figure~\ref{fig:nsbl_nsbb}.
The luminosity of the surface blackbody for the $7 \times10^6$ K and $10\times10^6$~K models is 0.8\% and 3.3\% of
the total luminosity respectively,
so its inclusion makes a difference in the best-fit continuum
parameters. This is most marked for the disc, as the shape of the
surface emission overlaps most with this component, so its inclusion
reduces the mass accretion rate and $L_{disc}/L_{BL}$. 

We use these models to determine the equivalent width of the iron
absorption lines from the surface against the 
brighter continuum emission from the accretion flow.
We used {\tt fakeit} command in XSPEC without
errors to produce a model spectrum on a dummy (diagonal) response matrix
from 1 keV to 15 keV with energy resolution of 1 eV.
We then fit this over the very restricted energy range of 4.7 keV to 5.7 keV
with a {\tt powerlaw + Gaussian + Gaussian} model. A single Gaussian
is a good representation of the multiple $n=2~\mathchar`-~1$ transitions in each ion
state as the substructure is blended due to the Doppler shifts (see
Section~\ref{sec:nsmodel}). Table~\ref{table:estimation_of_EW} shows the resulting
equivalent and intrinsic widths. The intrinsic width is higher for the
Fe XXV as there is a larger energy range between the multiple
transitions (forbidden, intercombination and resonance) than for
Fe XXVI (just spin-orbit splitting). Both lines increase in equivalent
width for higher temperature, as the contribution of the surface
emission increases, but Fe XXV always has higher equivalent width than
Fe XXVI, from $0.8~\mathchar`-~7.7$~eV as the temperature increases from $7~\mathchar`-~10
\times10^6$~K. 
In the next section, we search the higher resolution {\it Chandra} transmission grating data for the absorption lines,
and compare the result with these model predictions.

\begin{table}
\begin{center}
\begin{tabular}{cccc}
\hline \hline
& & \multicolumn{2}{c}{surface temperature} \\
Componet & Parameter & $7 \times10^6$ K & $10\times10^6$ K\\
\hline
{\tt kerrbb} 
& $M_\mathrm{dd}$ ($\mathrm{10^{18}\;g/s}$) & $0.620_{-0.010}^{+0.010}$ & $0.592_{-0.007}^{+0.008}$\\
& hd & $1.651_{-0.020}^{+0.020}$ & $1.601_{-0.015}^{+0.017}$\\
\hline
{\tt nthComp} 
& Gamma & $6.2_{-3.3}^{+0.3}$ & $6.4_{-3.3}^{+0.4}$\\
& $kT_\mathrm{e}$ (keV) & $> 3.0$ & $>  3.2$\\
& $kT_\mathrm{bb}$ (keV) & $1.75_{-0.15}^{+0.03}$ & $1.79_{-0.11}^{+0.03}$\\
& $L_\mathrm{disc} / L_\mathrm{BL}$ & $0.913_{-0.028}^{+0.029}$ & $0.870_{-0.020}^{+0.023}$\\
\hline
{\tt Gaussian}
& Line E (keV) &$6.657_{-0.032}^{+0.030}$ & $6.675_{-0.023}^{+0.023}$\\
& $\sigma$ (keV) & $0.233_{-0.036}^{+0.039}$ & $0.244_{-0.027}^{+0.029}$\\
& norm ($10^{-3}$)& $2.62_{-0.33}^{+0.35}$ & $3.43_{-0.33}^{+0.35}$\\
\hline
{\tt constant} (HXD PIN)
& & $1.208_{-0.044}^{+0.045}$ & $1.16_{-0.041}^{+0.043}$\\
\hline
$\chi^2/d.o.f.$ & & 2080.5 / 1918 & 2131.4 / 1918\\
\hline
\hline
\end{tabular}  
\caption[t]{Best-fit parameters for {\tt kerrbb + nthcomp} with the surface blackbody models using {\it Suzaku} spectrum. $N_\mathrm{H}$ is $4.4 \times 10^{21}~ \mathrm{cm^{-2}}$ (fixed).}
\label{table:nsbl_nsbb}
\end{center}
\end{table}

\begin{table*}
\begin{center}
\begin{tabular}{cc|ccc}
\hline \hline
& effective temperature & EW(eV) & Energy (keV) & $\sigma$ (eV)\\
\hline
Fe XXV x+y,w,z & $7 \times10^6$ K & 0.76 & 5.11 & 83\\
Fe XXVI Ly$\alpha$ &            & 0.05 & 5.35 & 39\\
\hline
Fe XXV x+y,w,z & $10 \times 10^6$ K & 7.7 & 5.12 & 86 \\
Fe XXVI Ly$\alpha$ &			 & 2.0 & 5.35 & 53 \\
\hline
\hline
\end{tabular}  
\caption[t]{Estimation of the equivalent width of the iron absorption line with different surface temperatures.}
\label{table:estimation_of_EW}
\end{center}
\end{table*}

\section{Line search with {\it Chandra}} 
\label{sec:chandra_data_analysis}
%==============================================
\subsection{Observations and data reduction}

{\it Chandra} observed Serpens X-1 with the High Energy Transmission
Grating Spectrometer (HETGS) twice on 2014 June 27 - 29 and August 25
- 26 (Obs. ID : 16208, 16209) each with $\sim 150$~ks exposure. These
data were taken in continuous clocking (CC) mode in which the data are
transferred every 2.85 ms in order to avoid pileup. The HETG consists
of two sets of gratings, the Medium Energy Grating (MEG) and the High
Energy Gratings (HEG) \citep{Canizares:2005}. The two gratings have
different grating periods, 4001.41 {\AA} in MEG and 2000.81 {\AA} in
HEG. Since the MEG's grating period is almost exactly twice as long
as the HEG's, the scattering angle for the N-th order HEG is very
close to that of 2N-th order MEG. The alignment angles for MEG and
HEG are very similar ($-5.19^\circ$ in HEG and $4.74^\circ$ in MEG),
so the position along the X-axis of the N-th order HEG and 2N-th order MEG are almost the same as each other.

In the standard (timed exposure: TE) mode, the full imaging is
retained, and the MEG and HEG grating orders are separated in Y-axis
values as well as X-axis and so do not overlap. However, in CC mode,
the information about the position in the columns (CHIPY) is lost
because events are read out continuously and the frame image is
collapsed into one row. Thus the N-th order HEG and 
2N-th order MEG overlap. Normally, in order to
avoid this, the source point is placed with a Y-axis offset 
from the center of the chip so that the MEG positive order and HEG
negative order (or MEG negative order and HEG positive order) are
excluded from the chips. However, in our data set, the source point
is set at the center of the Y axis, so MEG$\pm 2$ are intermixed with
HEG$\pm 1$ and the response files (which assume HEG and MEG are
separated) are not appropriate. Hence we use only
MEG$\pm$1 in our analysis. The MEG effective area is comparable to
that of the HEG at 5~keV, so this is not a big loss of signal (see Figure~\ref{fig:eff}). 

We processed the data using {\tt chandra\_repro} command in CIAO V4.8
software package and combined data from both observations to produce
spectra for MEG+1 and MEG-1 (there are no MEG$\pm 2$ data from this
tool due to the overlap with HEG$\pm 1$ discussed above). 
There is
one weak type I burst during the observations, but we did not exclude
it because its contributes only $\sim$0.5 \% of the total counts
\citep{Chiang:2016}. We fixed the hydrogen column density,
photoelectric absorption cross-sections and the metal abundances, as
for the {\it Suzaku} analysis.
 
\subsection{Method of absorption line search}
\label{sec:cal_3sigma}

\begin{figure}
\begin{center}
\includegraphics[bb = 0 0 612 792, width=60mm, angle=-90]{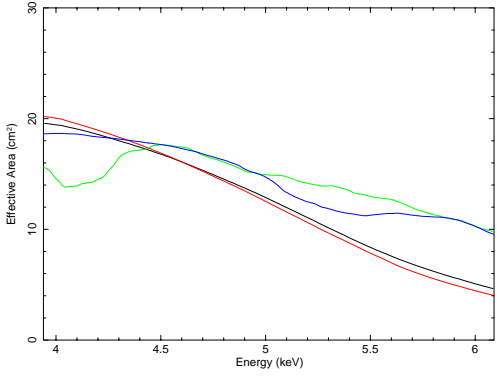}
\caption{{\it Chandra} HETG effective areas. The black curve is MEG-1, the red one is MEG+1, the green one is HEG-1 and the blue one is HEG+1.}
\label{fig:eff}
\end{center}
\end{figure}

We do a blind search on the {\it Chandra} MEG$\pm 1$ data. The target
of the search is the combination of Fe XXV x+y, w, z (6.7 keV) since
it is the strongest absorption feature in the neutron star atmosphere
model over the expected temperature range. We set the range of search
redshift as z = 0.1 $\sim$ 0.7, corresponding
to an energy for Fe XXV of 6.09 keV to 3.94 keV (though the most likely
range of neutron star radii is 10 - 15~km i.e. $z=0.31~\mathchar`-~0.17$). No
discrete structures are seen in the MEG$\pm 1$ effective areas in this
range (Figure~\ref{fig:eff}).

\begin{figure}
\begin{center}
\includegraphics[bb = 0 0 750 600,width=80mm]{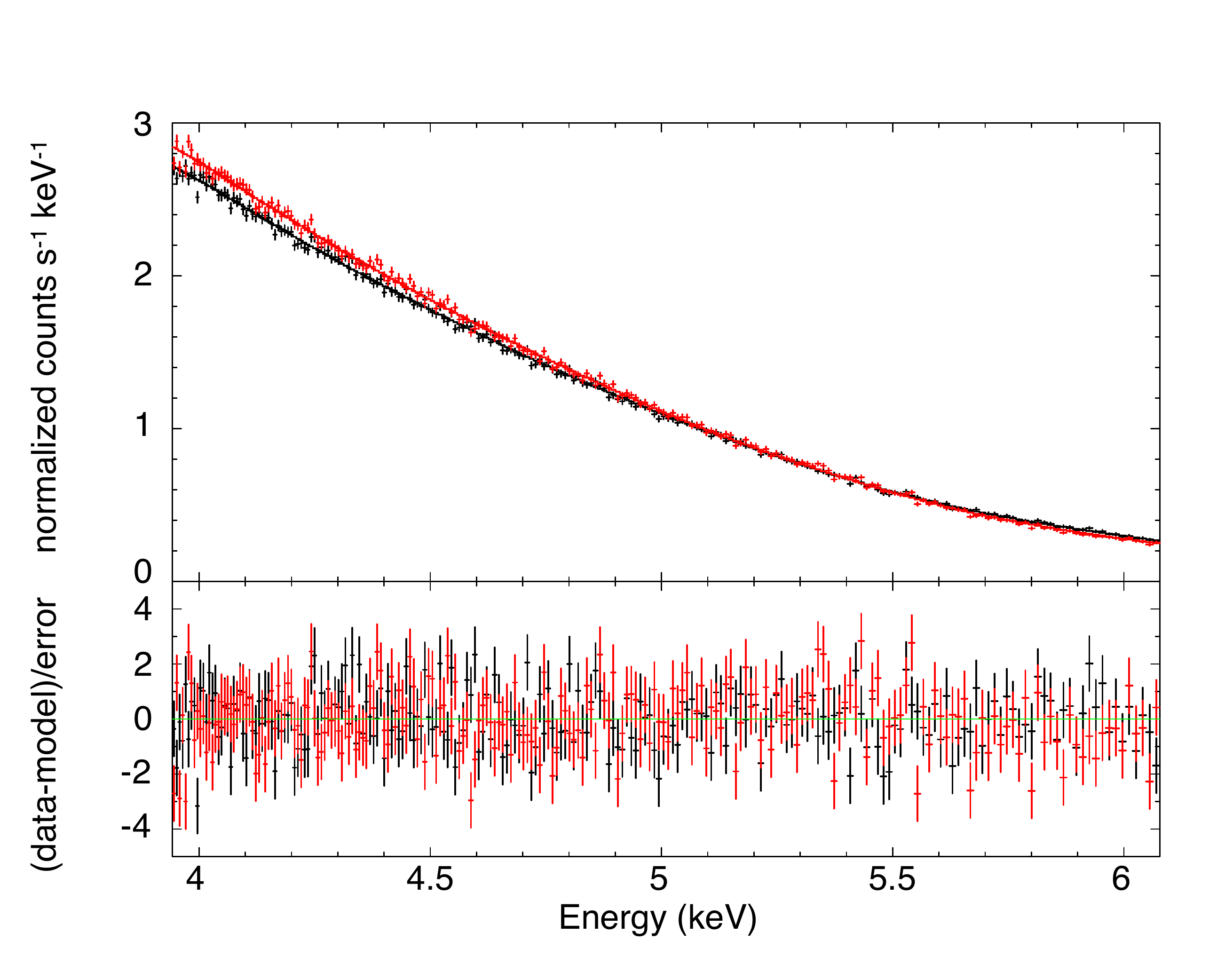}
\caption{{\it Chandra} MEG$\pm1$ count rate spectrum and the best-fit continuum model.
The black curve is MEG-1 and the red one is MEG+1.}
\label{fig:continuum_meg}
\end{center}
\end{figure}

We fit the MEG$\pm 1$ spectra separately with separate {\tt powerlaw}
models from 3.90 keV to 6.13 keV (i.e. extended by 0.04~keV from the
central energy range so that the line would be fully resolved by the
data even at the edge of the redshift range). This adequately
describes the continuum over this narrow energy range (see
Figure~\ref{fig:continuum_meg}). 
We save the parameters of the best-fit model and the chi-squared values ($\chi^2_0$).
The parameters are not tied between the two spectra as there are differences of a few percent in the flux.
This is considered to be due to the scattering halo effect or clocking background which are associated with CC-mode~\citep{ccmodedoc}.
They do not affect the structure of the absorption line because they make the continuous background.
Thus they are of no significance in this analysis.

We fix the power law continuum and add in a Gaussian line,
parameterized by the center energy ($E$), standard deviation
($\sigma$) and equivalent width (EW). We fix $\sigma$ at 80 eV (see
Table~\ref{table:estimation_of_EW})
and vary $E$ from 3.94 keV to 6.09 keV in steps of 0.01 keV 
(much smaller than the expected intrinsic width of the line of 0.08 keV).
Thus the only free parameter in the fitting is
the EW, which can be either negative or positive,
corresponding to absorption and emission lines, respectively. We
calculate the difference of the chi-squared values from $\chi^2_0$ and
save it ($\Delta \chi^2(E)$) together with the EW of the best-fit
model (EW($E$)).

We determine the $3\sigma$ confidence limit on the detection of a line
i.e. find the $\Delta\chi^2$ which corresponds to a probability of 0.0027 of getting a false positive.
This is not straightforward due to the multiple energies searched.
We approach this in two different ways.
Firstly, we simulate MEG$\pm$1 spectra from a simple {\tt powerlaw} continuum. 
We fit these with {\tt powerlaw} $+$ {\tt Gaussian} stepping the center energy of the {\tt Gaussian} over the same energy range as for the real data. 
We repeat this 10000 times and rank the resulting $\Delta\chi^2(E)$,
picking the 27th highest (as 10000 $\times$ 0.0027 is 27). 
This gives $\Delta\chi^2_{3\sigma} = 15.12$.

We confirm this using a more sophisticated statistical approach from high energy physics, where searching for a signal 
from a particle of unknown mass is a classic problem (the look elsewhere effect). 
We follow \cite{Gross:2010} who show that the relevant
$\chi^2$ value is that for which 
\begin{eqnarray*}
\begin{split}
P\left(\mathrm{max(\Delta \chi^2(E))} > \Delta\chi^2_{3\sigma} \right) = \\
P(\chi_1^2 > \Delta\chi^2_{3\sigma}) &+ \ensuremath{\langle N(c_0)\rangle} e^{-(c-c_0)/2}
\end{split}
\label{eq:3sigma_cal}
\end{eqnarray*}
where $\chi^2_1$ is a $\chi^2$ distribution with 1 degree of freedom, and $\ensuremath{\langle N(c_0)\rangle}$ is the number of 'upcrossings' between $\Delta \chi^2(E)$ and $\Delta \chi^2 = c_0$, that is, the number of $E$ in the search range which satisfies that 
$\Delta \chi^2(E) = c_0$ and $\Delta \chi^2(E - \epsilon) < c_0$ ($\epsilon$ is sufficiently small.).
We follow \cite{Gross:2010} and set $c_0=0.5$.

We use the same simulations as before and find $\ensuremath{\langle N(c_0)\rangle}=4.42$. 
This implies $\Delta \chi^2_{3\sigma}=15.37$, very similar to 
the standard simulation result.
Hereafter this value is used for the blind search because the first result has a relatively large Poisson error $\sim 0.3$. 

\subsection{Results of absorption line search}

\begin{figure}
\begin{center}
\includegraphics[bb = 0 0 1100 700,width=85mm]{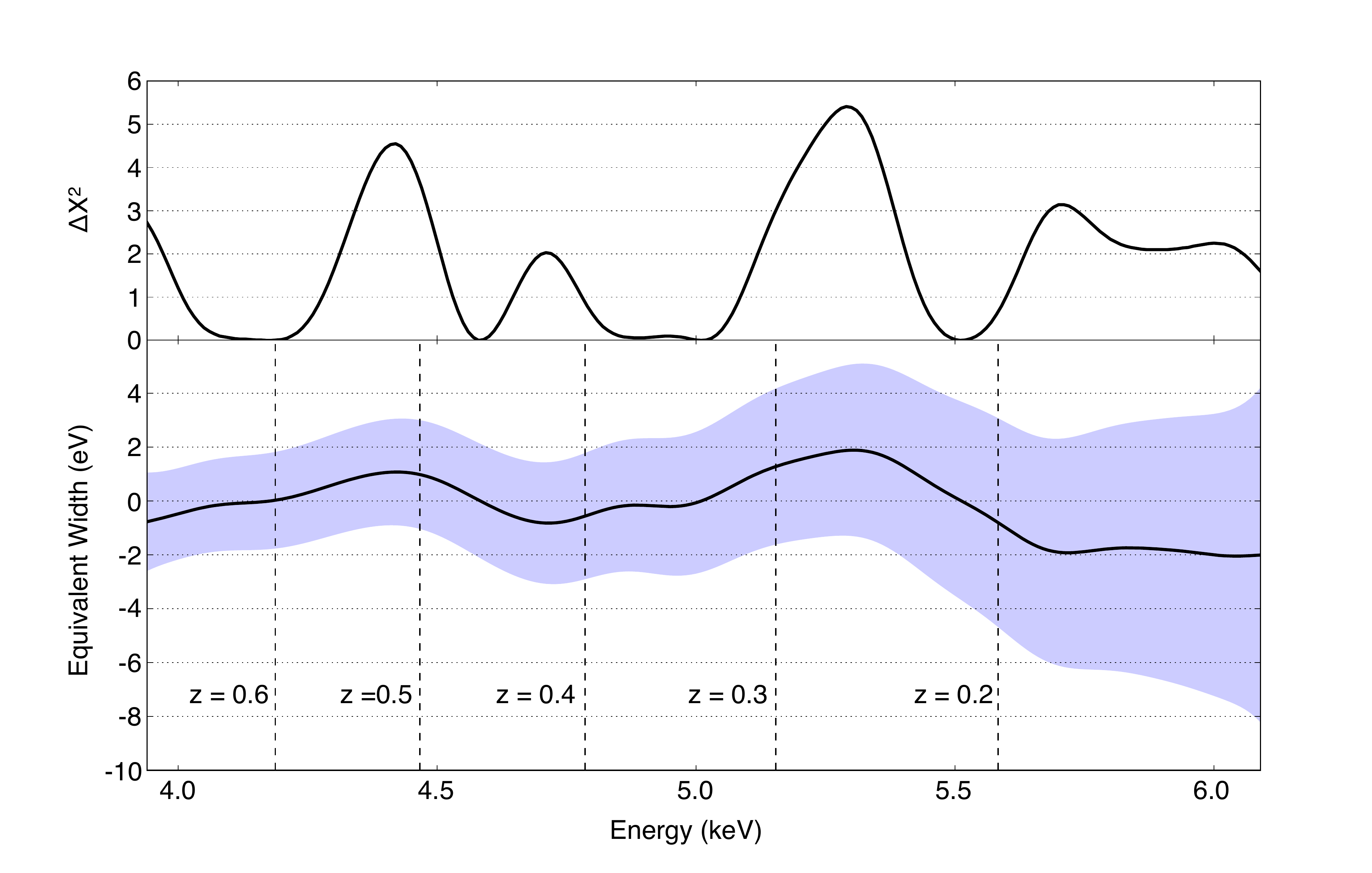}
\caption{Top: $\Delta \chi^2$ at different line energy in the search range. Bottom: Best-fit value of the equivalent width (the black solid line) and $3 \sigma$ upper limit of it.}
\label{fig:ew_meg}
\end{center}
\end{figure}

The maximum observed value of $\Delta \chi^2(E)$ is 5.41 (upper panel
in Figure~\ref{fig:ew_meg}). This is substantially smaller than the
value of 15.37 derived above for a $3\sigma$ detection. 
Therefore we did not find any absorption lines at the 3 $\sigma$ confidence level.
The $3\sigma$ upper limit of the equivalent width was found at each
energy by increasing the intensity until $\Delta \chi^2(E)=15.37$ was
reached. These are plotted in the lower panel of
Figure~\ref{fig:ew_meg}. 
The decrease in effective area at higher energies means that
this upper limit on any absorption line goes from $-8$~eV for $z=0.1$
to $\sim -2$~eV for $z\ge 0.3$. 
However, there are broad features present which probably indicate residual continuum curvature which is not well
modelled by the assumed power law. 
A more accurate measure of the upper limit of the line with respect to a more complex continuum is half the difference between the upper and lower limit to EW.
This gives $\sim -6, -4, -3, -1.5$ eV at z = 0.1, 0.2, 0.3, 0.7.

The upper limit from the {\it Chandra} data is $-3$~eV at $\sim 5.1$~keV.
This rules out an absorption line at $\sim 5.1$~keV with an EW as large as $7.7$~eV as predicted by the $10\times10^6$~K surface model. 
It is instead consistent with the predicted absorption line EW of 0.8 eV
for the lower surface temperature of $7\times10^6$~K.
We simulated this photosphere model through the {\it Chandra} response using the {\tt fakeit} command,
and estimate that such a line could be detected at the $3\sigma$ confidence level with a 3 - 4 Ms exposure.

%==============================================
\section{Discussion}
\label{sec:discussion}
%==============================================

The {\it Chandra} data clearly show that either the neutron star surface
temperature is lower than $10 \times 10^6$ K or some other mechanism suppresses
the absorption line. We first assess the expected surface temperature,
and then examine the other assumptions which determine the absorption
line strength. 

\subsection{Surface temperature}

We measured the surface temperature beneath the boundary layer as $\sim$ 1.7~keV
from the seed photon energy of the Comptonization component in the
broadband {\it Suzaku} data (Table~\ref{table:nsbl}). However the
temperature at the pole is not necessarily the same. The equator is 
illuminated by the boundary layer and subject to ram pressure heating, whereas
the pole is largely unaffected by these. Instead, the pole is heated
by thermal conduction from the equator and from the interior of the
star (though this is probably negligible in comparison). 

We calculate the distribution of the temperature on the surface solving the heat equation.
We assume the spherical crust of the neutron star with a thickness of $h$.
In spherical coordinates, it is described as:
\begin{eqnarray*}
\begin{split}
\rho C_p \frac{\partial T}{\partial t} = 
  \frac{1}{r^2}\frac{\partial}{\partial r} \left(\kappa r^2 \frac{\partial T}{\partial r}\right)
+ \frac{1}{r^2 \sin\theta}\frac{\partial}{\partial \theta} \left(\kappa \sin\theta\frac{\partial T}{\partial \theta}\right)\\
+ \frac{1}{r^2 \sin^2\theta}\frac{\partial}{\partial \phi} \left(\kappa \frac{\partial T}{\partial \phi}\right)
+ g
\end{split}
\end{eqnarray*}
$\rho$ is the mass density and $ C_p$ is the heat capacity and $\kappa$ is the thermal conductivity and $g$ represents external sources.
Assuming that $\frac{\partial T}{\partial t} = 0$ and $\kappa$ is constant in the crust, and ignoring $r$- and $\phi$- dependence for simplicity, the heat equation becomes: 
\begin{eqnarray*}
0 &=& \frac{\kappa}{R^2 \sin\theta}\frac{\mathrm{d}}{\mathrm{d} \theta} \left(\sin\theta\frac{\mathrm{d} T_\theta}{\mathrm{d} \theta}\right)
+ g\\
&=& \frac{\kappa}{R^2} \frac{\mathrm{d}}{\mathrm{d} \left(\cos \theta\right)} \left(\left(1-\cos^2\theta \right)\frac{\mathrm{d}}{\mathrm{d} \left(\cos \theta\right)}\right)T_\theta + g
\end{eqnarray*}
$R$ is the radius of the neutron star. 
When we integrate this equation along the $r$-axis from the inner crust ($r = R - h$) to the outer crust ($r = R$),
the external sources are a black body radiation from the surface and the heat transfer from the neutron star core.
Thus the heat equation in a spherical
shell at latitude $\theta$ 
(measured from the pole, so $\theta = \pi/2$ corresponds to the
equator) is 
\begin{eqnarray*}
  \sigma_s T_\theta^4 &=& \frac{\kappa h}{R^2} \frac{\mathrm{d}}{\mathrm{d} \left(\cos \theta\right)} \left(\left(1-\cos^2\theta \right)\frac{\mathrm{d}}{\mathrm{d} \left(\cos \theta\right)}\right)T_\theta + Q
\end{eqnarray*}
where $Q$ is the heat energy from the interior of the star.

\begin{figure}
\begin{center}
\includegraphics[bb = 0 0 567 292, width=80mm]{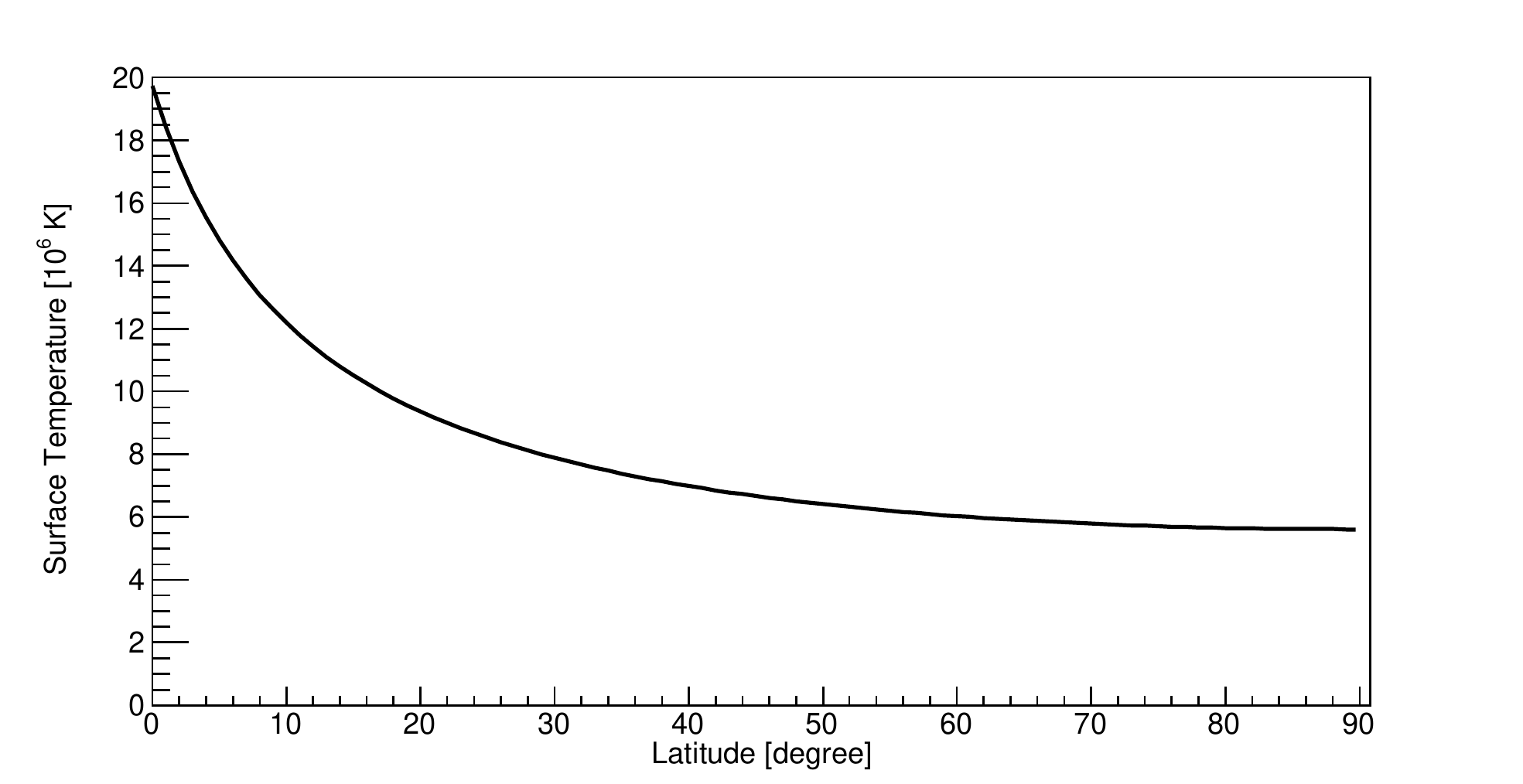}
\caption{Thermal distribution along the latitude. We set $T_0$ as $1.7$ keV and $\kappa$ as $10^{23}\;\mathrm{erg \;cm^{-1}\;K^{-1}\;s^{-1}}$.}
\label{fig:therm_dist}
\end{center}
\end{figure}

We assume $Q = 0$ for simplicity and set the boundary conditions as
$T_{\theta = \pi / 2}=T_0$ and 
\begin{eqnarray*}
\frac{R^2\sigma_s}{\kappa h} T_\theta^4 + 2\frac{\mathrm{d}T_\theta}{\mathrm{d}(\cos\theta)} = 0\;(\mathrm{at}~\theta = 0)
\end{eqnarray*}
The latter equation represents non-divergence of second-order
differentiation of $T$ at the pole. The thermal conductivity $\kappa$
in the crust is from $10^{18}$ to $10^{24}$ erg cm$^{-1}$ K$^{-1}$
s$^{-1}$ depending on the density and magnetic field
\citep{Geppert:2004}. Figure~\ref{fig:therm_dist} shows the thermal
distribution when we set $T_0$ and $\kappa$ to be 1.7 keV and
$10^{23}$ erg cm$^{-1}$ K$^{-1}$ s$^{-1}$. 
We assume $h=10^5~\mathrm{cm}$ and $R=10^6~\mathrm{cm}$.
This gives a temperature at
the pole of $\sim 5.6 \times 10^6$ K.

\begin{figure}
\begin{center}
\includegraphics[bb = 0 0 567 292, width=80mm]{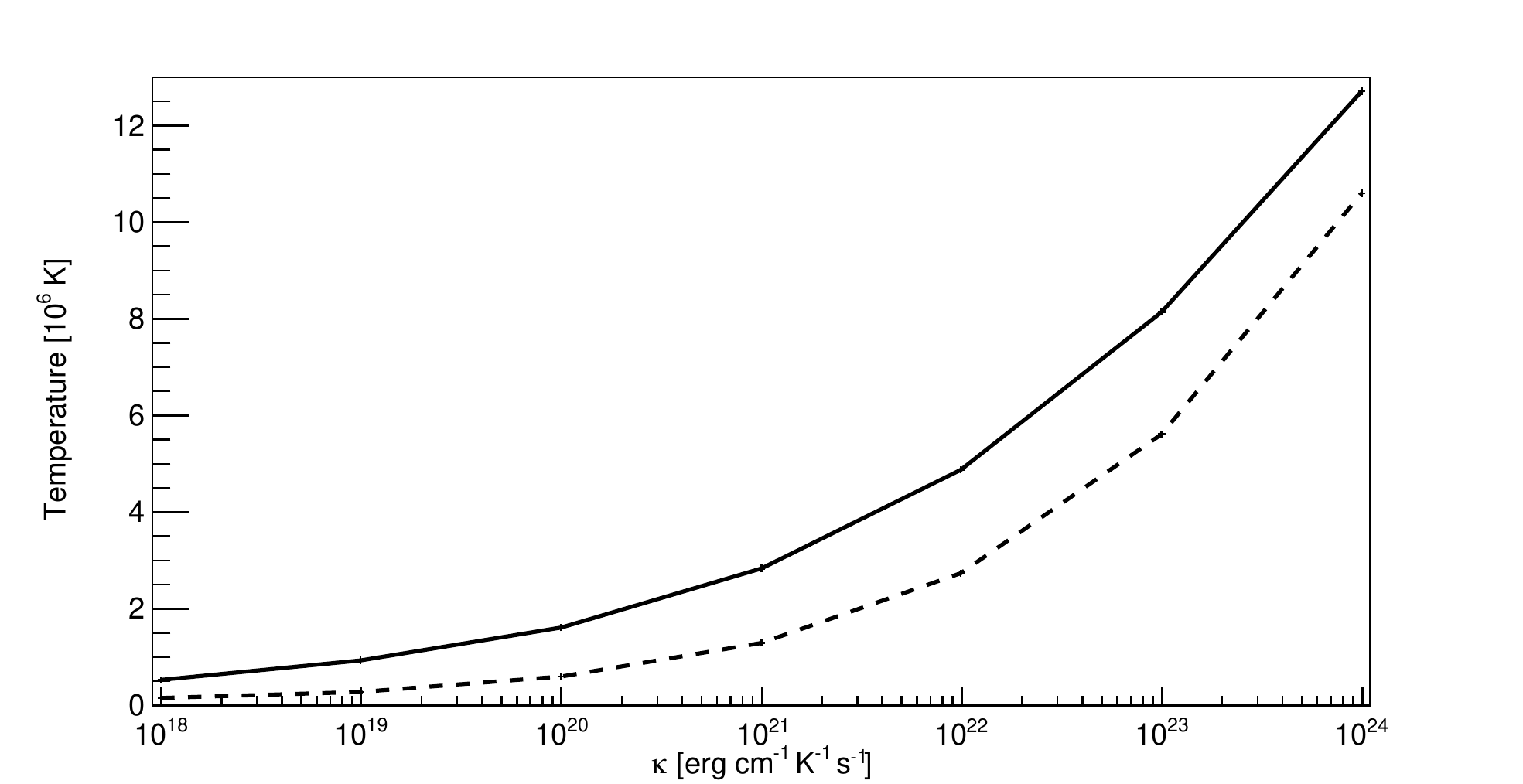}
\caption{Dependence of the surface temperature on $\kappa$ : The solid line is the temperature averaged over the surface, $T_\mathrm{ave}$. The dashed line is the temperature at the pole, $T_{\theta = 0}$.
The range of $\kappa$ is derived from \protect\cite{Geppert:2004}.}
\label{fig2}
\end{center}
\end{figure}

Observationally we see the angle averaged surface temperature.
We define $T_{\mathrm{ave}}$ defined by
\begin{eqnarray*}
\sigma_s T_{\mathrm{ave}}^4 \pi R^2 &=& \int_0^{\frac{\pi}{2}} \sigma_s T_\theta^4 2\pi R^2 \sin\theta \cos\theta \mathrm{d}\theta
\end{eqnarray*}
as the average of the blackbody radiation ($\sigma_s T_\theta^4$) over the surface.
It is $\sim 8\times10^6$ K for the model shown in Figure~\ref{fig:therm_dist}. This
temperature is close to that of the lower temperature simulation, so
predicts a line EW of a few eV, consistent with the limit from the
{\it Chandra} data. However, it is more likely that $\kappa$ is
lower. \cite{Rutledge:2002} discuss the crustal conductivity
calculated from from electron-ion (appropriate for the envelope) 
and electron-phonon scattering (as appropriate for the crystalline
phase, see also \citealt{Potekhin:2015}). 
This gives $\kappa=10^{19-20}$ erg cm$^{-1}$ K$^{-1}$ s$^{-1}$, consistent with
observations of neutron star cooling after a transient accretion
episode \citep{Rutledge:2002}. Figure~\ref{fig2} shows the dependence of $T_{\mathrm{ave}}$ on
$\kappa$. For such low values of the thermal conductivity, the 
angle averaged temperature is $\sim 1~\mathchar`-~2\times10^6$ K and 
a very little absorption line is produced. 
In addition, the distribution of the iron ionization state becomes so complex that the line identification is too complicated even if it is observed (Figure~\ref{fig:ion}). 

These low temperatures are very close to the temperatures observed for
the neutron star surface cooling after a transient accretion episode
\citep{Heinke:2013, Degenaar:2015}. This emission is powered by deep
crustal heating by pycnonuclear reactions during accretion, and sets a
lower limit to the temperature of the pole even when thermal
conductivity is low. This points to the need for more sophisticated
calculations which include the energy exchanged with the inner part of
the star. Nonetheless, it appears feasible that the surface
temperature of the neutron star surface near the pole is less than
$7\times 10^6$~K and so produces an Fe XXV absorption line EW which is
less than 1~eV, undetectable with current data. 
However, there are other possible issues which could suppress the line EW.
We explore them below.

\subsection{Uncovered surface area and metallicity}

The disc accretion onto the neutron star surface terminates in a boundary
layer, where the deceleration takes place in an equatorial belt
with meridional extent set by the mass accretion rate \citep{Inogamov:1999}.
The boundary layer is optically thick, so shields the
surface from view, and is hot enough that iron should be completely
ionized so it produces no atomic features. The extent of the boundary
layer depends on mass accretion rate, and for the
$0.3~\mathchar`-~0.6~L_\mathrm{Edd}$ (distance of 7.7 - 10~kpc) derived for Serpens X-1 in Section~\ref{sec:suzaku_data_analysis}, the accretion
belt will cover around half of the visible neutron star surface.
This will halve the predicted line EW for a given temperature, but 
a surface as hot as $10^7$ K would still be marginally inconsistent
with the {\it Chandra} upper limits for a 10-13~km neutron star. 
However, the geometry of the boundary layer also depends on other physical parameters.
For example, in the case that the disc thickness on the neutron star surface is relatively thick,
the boundary layer can reach the pole if $L_\mathrm{BL} > 0.25~\mathchar`-~0.4~L_\mathrm{Edd}$ \citep{Suleimanov:2006}.
Thus we cannot rule out the possibility that the neutron star surface is totally covered by the optically thick corona.

The extent of the boundary layer also determines the metallicity of
the photosphere at the pole. Heavy metals sink on a timescale of
$\sim 10$~s due to the strong gravity of the neutron star
\citep{Bildsten:2003}, but even low levels of accretion are sufficient
to replenish iron in the photosphere \citep{Ozel:2013}. 
The meridional structure of the boundary layer means that 
the deposition of fresh material takes place in the equatorial belt rather than at the pole.
Hence it is possible that
there is an abundance gradient with polar angle, 
with the pole being mostly Hydrogen while the metals are confined to the hot boundary layer.
This anti-correlation of metallicity with 'bare' (uncovered by the
boundary layer) neutron star surface would mean that no iron
absorption lines could be seen whatever the polar temperature.
However, the amount of accretion required to replenish the photosphere
is very small, and the maximum polar angle extent of the boundary
layer is not sharp. There is a 'dark layer' of material beyond the
bright boundary layer which extends closer towards the poles,
especially for the fairly high accretion rates considered here
 \citep{Inogamov:1999}, so it seems unlikely that metals are
absent from the surface.

\subsection{Boundary layer illumination of atmosphere}

The model atmosphere calculations in Section~\ref{sec:nsmodel} assume that there is no
additional heating from illumination. Plainly the boundary layer will
illuminate the surface at the pole to some extent, though the
equatorial belt geometry means that only a small fraction of the
boundary layer flux will be intercepted by the polar regions. While
this could contribute to heating the surface, increasing the expected
absorption line depth, irradiation can produce a temperature
inversion which switches the line into emission. There are no current
calculations of this effect, but we note that our {\it Chandra} upper limits
are equally stringent for emission as for absorption. 

\subsection{Inclination offset and spin frequency}

The binary inclination is small, $\sim 10^\circ$
\citep{Cornelisse:2013}, but the neutron star spin axis could be
misaligned from the binary orbit due to the supernovae kick at its
formation \citep{Brandt:1995}. This leads to both
misalignment of the spin and orbit, and high orbital eccentricity but
both of these are removed by tidal forces, often before the companion
fills its Roche lobe and the system becomes an LMXB \citep{Hut:1981}. Thus
the neutron star spin in Serpens X-1 should be aligned with the binary
orbit, and the binary orbit inclination is low \citep{Cornelisse:2013}. 
However, measurements of the inclination from sophisticated
fitting of the reflected signature from the boundary layer
illumination of the accretion disc gives a significantly larger value,
$i\sim 27^\circ$ \citep{Matranga:2017}, which would be sufficient to
broaden the line beyond detectability as a narrow feature. Inclination
is difficult to measure precisely by reflection, as it depends on the
spectral modelling (compare to \citealt{Miller:2013} who used less physical
models but derived $i\sim 9^\circ$ from the same data). We conclude
that the inclination angle to the neutron star spin axis is most
likely low, so this is not the origin of the loss of narrow line. 

The spin frequency of Serpens X-1 has not been detected.
In our model prediction, we assume that it is 400 Hz.
This is a typical value in observed LMXBs \citep{Altamirano:2012, Patruno:2012}.
In the spectral analysis in Section~\ref{subsec:spectralcomponent},
the spin frequency is estimated as $\sim$ 500 Hz using 
the luminosity ratio of the disc emission to the boundary layer
emission, and it is consistent with that of LMXBs.
However if this object is a peculiar source and the spin frequency is over than $\sim$ 1 kHz,
the absorption lines would be totally broadened.

%==============================================
\section{Conclusions}
\label{sec:conclusions}
%==============================================

We show that atomic features from highly ionized iron are potentially
observable in the persistent (non burst) emission of accreting LMXB
for low inclination or low spin neutron stars. Serpens X-1 is the only
known non-transient system which fulfils these constraints, having low
orbital inclination. 
This is on the middle banana branch, 
so the accretion flow geometry is probably an accretion disc
which forms a boundary layer around the neutron star equator,
leaving the polar surface directly visible.
We use the {\it Suzaku} broadband data to model the continuum emission, 
and then use this as a baseline to add in the predicted surface emission. 
We model this for two temperatures which span a reasonable range for the polar surface
of $7~\mathchar`-~10\times10^6$ K. 
These predict an absorption line from Fe XXV K$\alpha$ with equivalent width of 0.8 - 8 eV for a completely  equatorial boundary layer,
or 0.4 - 4 eV if the boundary layer covers half of the neutron star surface. 
We search for this line in existing {\it Chandra} grating data, 
and find an upper limit of 2 - 3 eV. 
We discuss potential reasons for this non-detection 
(the surface temperature, the geometry of the boundary layer,
the metallicity in the atmosphere, the inclination angle and the spin frequency of the star). 
Our conclusion is that the line is likely there at the level of 1eV, 
a combination of the boundary layer obscuring half of the surface,
and the polar temperature being lower than $10\times 10^6$~K. 
However, to detect such a line at $3\sigma$ confidence would require 3-4 Ms exposure time with the {\it Chandra} gratings. 
Thus it is unlikely that current instrumentation can obtain substantially better constraints.
The effective area of {\it XARM} (the {\it Hitomi} recovery mission) is $\sim$ 10 times as large as that of {\it Chandra} around 5 keV,
so the line could be detected at $3\sigma$ confidence in $\sim$ 300 ks.
The future X-ray mission {\it Athena} has a much larger effective area (Figure 4 in  \citealt{Barret:2013}), so should be able to constrain such a small EW with an observation of around $2\times10^4$~s. 
The potential of such future observations motivates more theoretical work on the neutron star surface emission
(including a two dimensional analysis of the temperature, abundances and effect of illumination)
in order to obtain a better understanding of the expected features.

%==============================================
\section*{Acknowledgements}
%==============================================
The authors would like to express their thanks to V. Suleimanov and M. Baub\"{o}ck for making their available to us, and K. Ishibashi for advice on {\it Chandra} data analysis.
H.Y. acknowledges the support of the Advanced Leading graduate course for Photon Science (ALPS).
C.D. acknowledges the support from STFC under grant ST/L00075X/1 and a JSPS long-term fellowship.
This work was supported by the Grant-in-Aid for Scientific Research on Innovative Areas "Nuclear Matter in Neutron Stars Investigated by Experiments and Astronomical Observations" (KAKENHI 24105007).

\appendix
%==============================================
\section{MEG$\pm$1 and HEG$\pm$1}
\label{sec:MEG+HEG}
\begin{figure}
\begin{center}
\includegraphics[bb = 0 0 750 600,width=80mm]{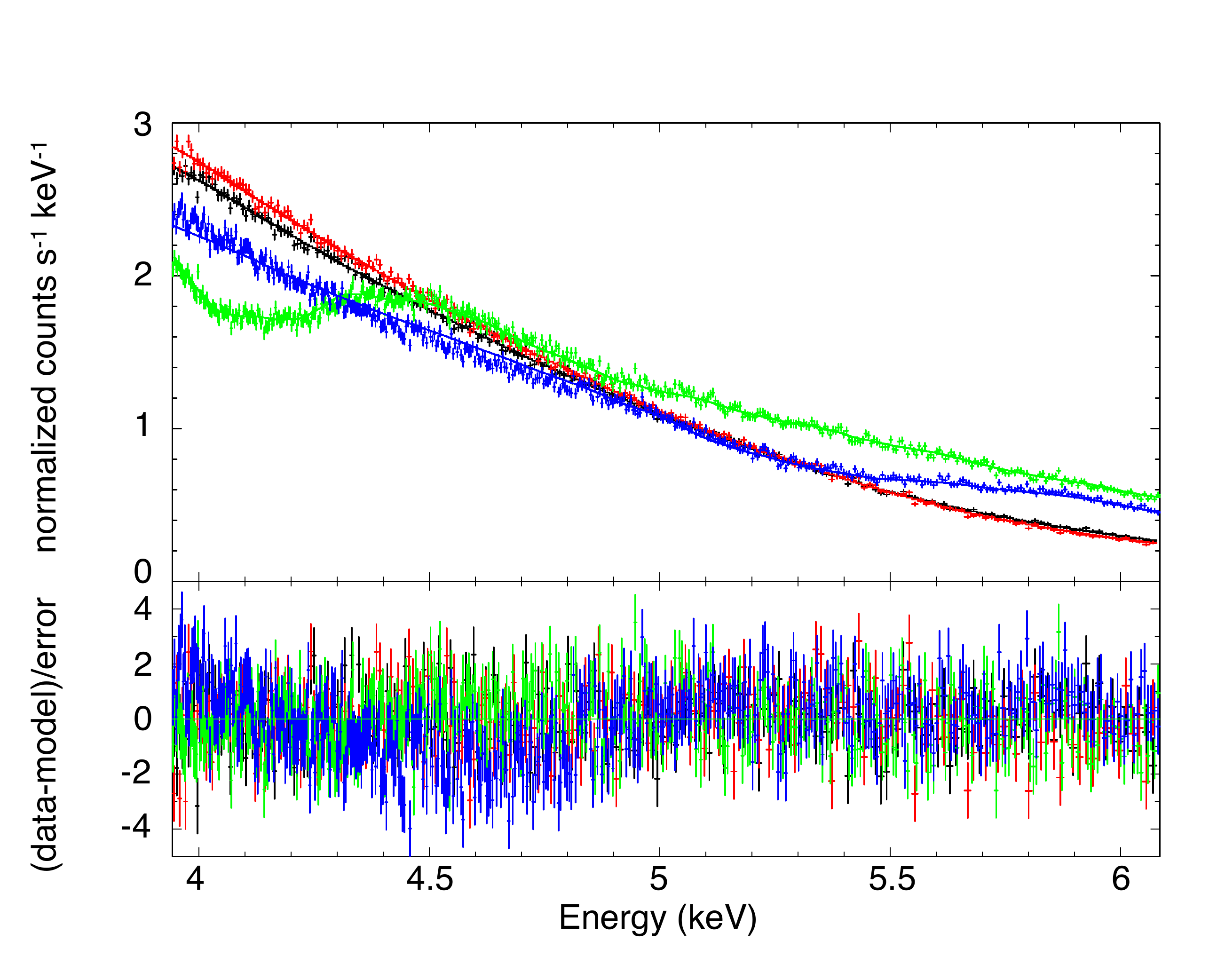}
\caption{{\it Chandra} MEG$\pm1$ \& HEG$\pm$1 count rate spectrum and the best-fit continuum model.
The black curve is MEG-1, the red one is MEG+1, the green one is HEG-1 and the blue one is HEG+1.}
\label{fig:continuum_all}
\end{center}
\end{figure}

\begin{figure}
\begin{center}
\includegraphics[bb = 0 0 1100 700,width=85mm]{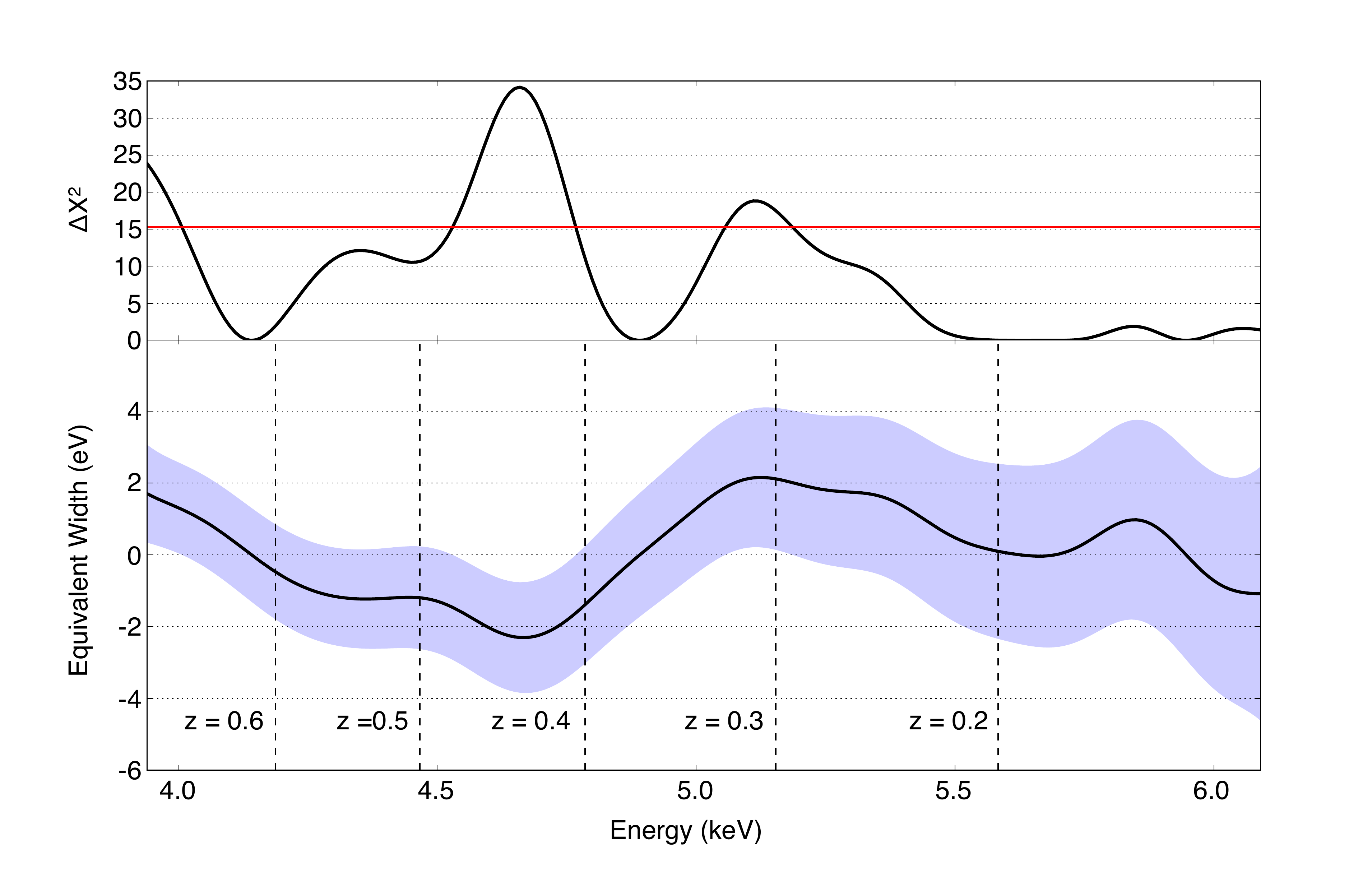}
\caption{Top: $\Delta \chi^2$ at different line energy in the search range. Bottom: Best-fit value of the equivalent width (the black solid line) and $3 \sigma$ upper limit of it. The red line represents for 3 $\sigma$ detection level (${\Delta \chi^2_\mathrm{3\sigma}} = 15.27$).}
\label{fig:ew_all}
\end{center}
\end{figure}

%==============================================
Just for reference, we show the result in Figure~\ref{fig:continuum_all}, \ref{fig:ew_all} when HEG$\pm$1 data are included to the analysis.
We obtained $\ensuremath{\langle N(0.5)\rangle} = 4.20$ and ${\Delta \chi^2_\mathrm{3\sigma}} = 15.27$.
(The number of trials of Section~\ref{sec:cal_3sigma} (iii) is 1000.)
As shown in Figure~\ref{fig:continuum_all}, the HEG+1 spectrum is rolling and the {\tt powerlaw} model does not fit it well.
Due to this, a large $\Delta \chi^2$ is seen in Figure~\ref{fig:ew_all} around 4.7 keV where the residuals is significant in HEG+1.
{\it Therefore the $\Delta \chi^2$ around 4.7 keV is larger than $\Delta \chi^2_\mathrm{3\sigma}$ but this does not mean that we have detected the absorption line.}
Considering that the other spectra are fitted well simply with the {\tt powerlaw} model, 
it is natural that the spectrum data or the response file are not correct due to the mixing.
If we use the response file which consider that the HEG+1 spectrum is intermixed with the MEG+2, the matters might be improved.

\bibliographystyle{mnras}
\bibliography{papers,RelativisticLineOfSerx1}

\end{document}